\documentclass[nofootinbib,aps,prb,preprint,notitlepage,superscriptaddress,onecolumn,longbibliography]{revtex4-1}
\usepackage{hyperref}
\hypersetup{colorlinks=true,
		    linkcolor=blue,
		    citecolor=blue,
		    allcolors=blue}
\usepackage{natbib}
\usepackage{amsmath, amsfonts, bm, bbm, braket, mathtools}
\usepackage{graphicx,overpic}
\usepackage{textcomp}
\graphicspath{{./}}

\usepackage{lineno}

\begin{document}


\title{Strategy to Extract Kitaev Interaction using Symmetry in Honeycomb Mott Insulators}
\author{Jiefu Cen}
\affiliation{Department of Physics, University of Toronto, Toronto, Ontario, Canada M5S 1A7}
\author{Hae-Young Kee}
\email[]{hykee@physics.utoronto.ca}
\affiliation{Department of Physics, University of Toronto, Toronto, Ontario, Canada M5S 1A7}
\affiliation{Canadian Institute for Advanced Research, CIFAR Program in Quantum Materials, Toronto, Ontario, Canada, M5G 1M1}
\date{\today}

\maketitle

\vspace{-2mm}\section*{\hspace{-7mm} Abstract}\vspace{-2mm}
{\bf The Kitaev spin liquid, a ground state of the bond-dependent Kitaev model in a honeycomb lattice has been a centre of attraction, since a microscopic theory to realize such an interaction in solid-state materials was discovered. A challenge in real materials though is the presence of the Heisenberg and another bond-dependent Gamma interactions detrimental to the Kitaev spin liquid, and there have been many debates on their relative strengths. Here we offer a strategy to extract the Kitaev interaction out of a full microscopic model by utilizing the symmetries of the Hamiltonian. Two tilted magnetic field directions related by a two-fold rotational symmetry generate distinct spin excitations originating from a specific combination of the Kitaev and Gamma interactions. Together with the in- and out-of-plane magnetic anisotropy, one can determine the Kitaev and Gamma interactions separately.  Dynamic spin structure factors are presented to motivate future experiments. The proposed setups will advance the search for Kitaev materials.}

\vspace{-2mm}\section*{\hspace{-7mm} Introduction}\vspace{-2mm}

An electron's orbital motion in an atom generates a magnetic field which influences its spin moment, known as spin-orbit coupling. 
When the coupling is strong in heavy atoms, the effective Hamiltonian is described by the spin-orbit-entangled pseudospin wave-function and
the interactions among magnetic ions are highly anisotropic different from the standard Heisenberg interaction \cite{balents2014review,rau2016review,winter2017review,hermanns2018review,Janssen2019review,Takayama2021JPSJ}.
A fascinating example is the Kitaev model with a bond-dependent interaction in a two-dimensional honeycomb lattice, whose ground state is
a quantum spin liquid (QSL) with Majorana fermions and Z$_2$ vortex excitations \cite{kitaev2006}.
There have been extensive studies on the model because in the Kitaev QSL
non-Abelian excitations emerge under a magnetic field, and their braidings provide topological computation.
Since a microscopic mechanism to generate such an interaction was uncovered\cite{jk2009prl},
intense efforts toward finding QSLs including 
a variety of candidate materials from spin $S$ =1/2\cite{singh2012relevance,choi2012prl,plumb2014prb,modic2014realization,HSKim2015prb,sears2015prb,sandilands2015continuum,johnson2015monoclinic,banerjee2016proximate,HSKim2016structure} to higher-spin $S$  systems have been made \cite{Peter2019HigherK,Lado2017XXZ,Xu2018Kitaev,Lee2020PRL}.
Despite such efforts, a confirmed Kitaev QSL is still missing.

One challenge in finding the Kitaev QSL in magnetic materials is the presence of other spin interactions which may generate 
magnetic orderings or other disordered phases \cite{cjk2010prl,cjk2014zigzag,rau2014prl,rau2014trigonal,winter2016challenges,janssen2017model,Luo2021npj}.
A generic nearest neighbour (n.n.) model in an ideal honeycomb was derived which revealed
the isotropic Heisenberg interaction and another bond-dependent interaction named the Gamma ($\Gamma$) \cite{rau2014prl}.
Furthermore, there exist further neighbour interactions such as second and third n.n. Heisenberg interactions,
which makes it difficult to
single out the Kitaev interaction itself.
There have been many debates on the relative strengths, especially between the dominant Kitaev and Gamma interactions in
Kitaev candidate materials\cite{HSKim2015prb,HSKim2016structure,janssen2017model,MaksimovPRR2020}, and an experimental guide on how to extract
the Kitaev interaction out of a full Hamiltonian is highly desirable.

In this work, we present a symmetry-based experimental strategy to determine the Kitaev interaction. 
Our proposal is based on the $\pi$-rotation around the $a$-axis perpendicular to one of the bonds in the honeycomb plane, denoted by $C_{2a}$ symmetry
that is broken by a specific combination of the Kitaev and $\Gamma$ interactions. 
This broken $C_{2a}$ can be easily detected with the help of a magnetic field applied within the $a-c$ plane where the $c$-axis is perpendicular to the honeycomb plane;
spin excitations under the two field angles of $\theta$ and $-\theta$,  measured away from the honeycomb plane 
as shown in Fig. \ref{Fig1_lattice}(a),  are distinct due to the combination of the Kitaev and Gamma interactions.
The two field angles are related by the $\pi$-rotation around $a$-axis, i.e. $C_{2a}$ operation.
Such differences are based on the symmetry and signal the relative strengths of these interactions.
A magnetic ordering that further enhances the broken $C_{2a}$ symmetry does not alter the asymmetry,  but quantifying
the interaction strengths requires the size of the magnetic ordering. 
For this reason, a polarized state in the high-field region would be ideal for our purpose. 

To determine each of the interactions, one needs to use the conventional in- vs. out-of-plane anisotropy in spin excitations.
We note that the Gamma interaction affects the conventional anisotropy, but the Kitaev does not when the field is large enough to compensate the order by disorder effect\cite{GiniyatPRB2016}.
 Thus subtracting the Gamma contribution deduced from the conventional anisotropy 
 allows us to estimate the Kitaev interaction from the measured spin excitations under the field angles of $\theta$ and $-\theta$. 
Both the conventional anisotropy and the $\pi$-rotation-related spin excitations 
can be  measured by angle-dependent ferromagnetic resonance (FMR) or inelastic neutron scattering (INS) techniques while sweeping the magnetic field directions in the $a-c$ plane containing the $C_{2a}$ rotation axis.

Below we present the microscopic model and main results based on the $\pi$-rotation symmetry around $a$-axis. 
To demonstrate our theory, we also show the FMR and dynamical spin structure factors (DSSF) obtained by exact diagonalization (ED). 
We analyze the different spin excitations under the two field angles at finite momenta using the linear spin wave theory (LSWT), which further confirms our results based on the symmetry argument.
Our results will guide a future search of Kitaev materials. 

\vspace{-2mm}\section*{\hspace{-7mm} Result}\vspace{-2mm}

\begin{figure} 
\includegraphics[width=1.0\linewidth,trim={0mm 00mm 0mm 00mm}]{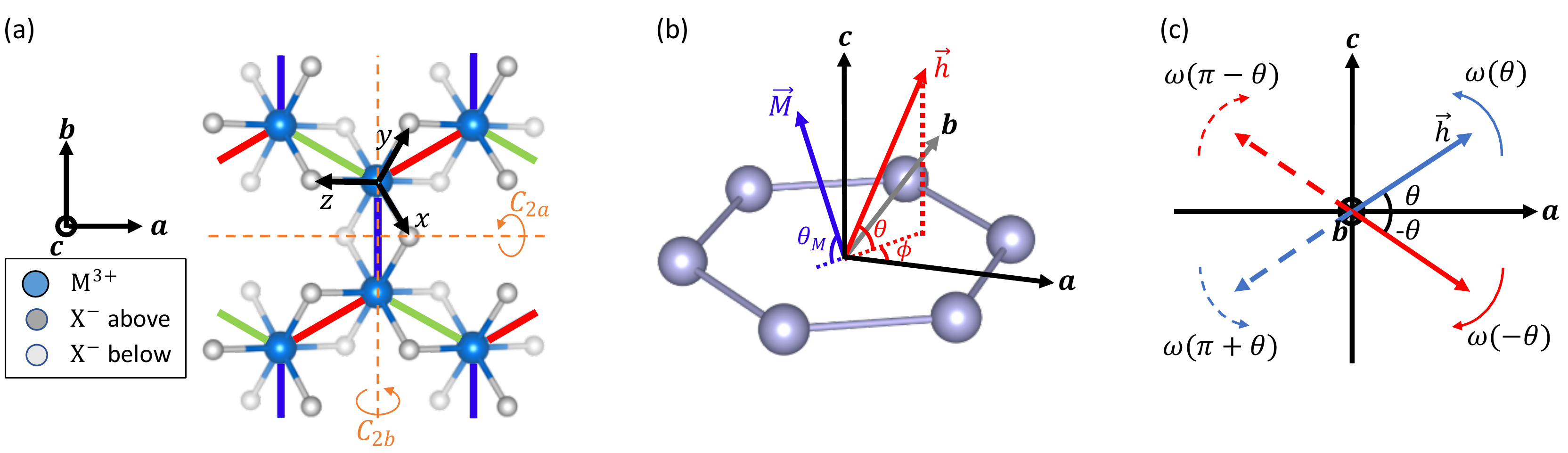}
\caption{\textbf{Crystal structure and direction of the magnetic field.} (a) Schematic of the honeycomb lattice of transition metal ions (light blue) in edge sharing octahedra environment of anions (above the honeycomb plane: gray,  below the plane: light gray).  Octahedral $\boldsymbol{x}\boldsymbol{y}\boldsymbol{z}$ axes,  $\boldsymbol{a}\boldsymbol{b}\boldsymbol{c}$ axes, and the Kitaev bonds x (red), y (green), z (blue) are indicated. $C_{2a}$ and $C_{2b}$ symmetries (orange) are highlighted. The octahedra environment breaks $C_{2a}$, while $C_{2b}$ symmetry is intact. 
(b) Direction of the external magnetic field $\vec{h}$ in $\boldsymbol{a}\boldsymbol{b}\boldsymbol{c}$ axes where $\theta$ is measured from the $a-b$ plane,  and $\phi$ is from the $a$-axis.  The blue arrow $\vec{M}$ represents the magnetic moment direction with the angle $\theta_M$.  (c) $\delta\omega_{K} (\theta)$ in the $a-c$ plane is the difference in the spin excitation energies $\omega$ between two field directions: $\omega(\theta)$ (blue) and $\omega(-\theta)$ (red). $C_{2b}$ maps $\omega (\theta)$ to $\omega (\pi+\theta)$, so $\delta\omega_K(\pi-\theta)=-\delta\omega_K(\theta)$.}
\label{Fig1_lattice}
 \end{figure}

\noindent {\bf Model} --
The  generic spin exchange Hamiltonian among magnetic sites with strong spin-orbit coupling for the ideal edge sharing octahedra environment in the octahedral $\boldsymbol{x}-\boldsymbol{y}-\boldsymbol{z}$ axes shown in Fig. \ref{Fig1_lattice}(a)
contains the Kitaev ($K$), Gamma ($\Gamma$), and Heisenberg ($J$) interactions\cite{rau2014prl}:
\begin{equation}
\mathcal{H}=\sum_{\langle ij\rangle\in\alpha\beta(\gamma)} \Big[ J{\bf S}_{i}\cdot {\bf S}_{j}+KS_{i}^{\gamma}S_{j}^{\gamma}+\Gamma(S_{i}^{\alpha}S_{j}^{\beta}+S_{i}^{\beta}S_{j}^{\alpha}) \Big],
\end{equation}
where ${\bf S} = \frac{1}{2} {\vec \sigma}$ with $\hbar \equiv 1$ and ${\vec \sigma}$ is Pauli matrix, 
 $\langle ij\rangle$ denotes the nearest neighbor (n.n.) magnetic sites, and $\alpha\beta(\gamma)$ denotes the $\gamma$ bond taking the $\alpha$ and $\beta$ spin components ($\alpha,\beta,\gamma\in\{\text{x,y,z}\}$). The x-, y-, and z-bonds are shown in red, blue, and green colours, respectively in Fig. \ref{Fig1_lattice}(a).
Further neighbour interactions and trigonal-distortion allowed interactions, and their effects will be discussed later. 

To analyze the symmetry of the Hamiltonian, we rewrite the model in the $\boldsymbol{a}-\boldsymbol{b}-\boldsymbol{c}$ axes \cite{Onoda2011,Ross2011JcpJpp,Chaloupka2015hidden}:
\begin{equation}\label{eq:hamiltonian}\begin{split}
\mathcal{H}&=\sum_{\langle i,j\rangle}\Bigg[J_{XY}(S_{i}^{a}S_{j}^{a}+S_{i}^{b}S_{j}^{b})+J_{Z}S_{i}^{c}S_{j}^{c}\\&\qquad+J_{ab}\left[\cos\phi_{\gamma}(S_{i}^{a}S_{j}^{a}-S_{i}^{b}S_{j}^{b})-\sin\phi_{\gamma}(S_{i}^{a}S_{j}^{b}+S_{i}^{b}S_{j}^{a})\right]\\&\qquad-\sqrt{2}J_{ac}\left[\cos\phi_{\gamma}(S_{i}^{a}S_{j}^{c}+S_{i}^{c}S_{j}^{a})+\sin\phi_{\gamma}(S_{i}^{b}S_{j}^{c}+S_{i}^{c}S_{j}^{b})\right]\Bigg], 
\end{split}\end{equation}
where
$\phi_{\gamma}=0,\frac{2\pi}{3}$, and $\frac{4\pi}{3}$ for $\text{\ensuremath{\gamma=\text{z-, x-, and y}}}$-bond respectively, and the exchange interactions are given by
\begin{equation}\begin{split}
J_{XY}&=J+J_{ac},\;\; J_{Z}=J+J_{ab},
\\J_{ab}&=\frac{1}{3}K+\frac{2}{3}\Gamma, \;\; J_{ac}=\frac{1}{3}K-\frac{1}{3}\Gamma.
\end{split}\end{equation}
The Hamiltonian $\mathcal{H}$ is invariant under $\pi$-rotation around the b-axis denoted by $C_{2b}$ and $\frac{2\pi}{3}$-rotation around the c-axis by $C_{3c}$ 
in addition to the inversion and time-reversal symmetry.

Our proposed experimental design is based on the observation that the $\mathcal{H}$ is {\it not} invariant under $\pi$-rotation about the a-axis $C_{2a}$
due to the presence of only $J_{ac}$, i.e., if $J_{ac} = 0$, $C_{2a}$ is also a symmetry of $\mathcal{H}$.
Since the $C_{2a}$ is broken by $J_{ac}$, if there is a way to detect the broken $C_{2a}$, that will signal the strength of $J_{ac}$. 
We note that the magnetic field sweeping from the c-axis to a-axis within the $a-c$-plane does the job.
The fields with angles of $\theta$ (blue line) and $-\theta$ (red line) for $ 0 < \theta < \frac{\pi}{2}$ shown in Fig. \ref{Fig1_lattice}(b) and (c) are related by $C_{2a}$ rotation, and 
thus measuring
the spin excitation difference between these two field directions will detect the strength of $J_{ac}$. 

To prove our symmetry argument, we consider a full model with a magnetic field. 
Under a magnetic field, the total Hamiltonian including the Zeeman term is given by
\begin{equation}
\mathcal{H}_{\text{tot}}=\mathcal{H}+\mathcal{H}_{B}=\mathcal{H}- g \; \mu_B \sum_{i}\vec{S}_{i}\cdot\vec{h},
\end{equation}
where the external field $\vec{h}$ has the polar angle $\theta$ measured away from the $a-b$ honeycomb plane and the azimuthal angle $\phi$ from the a-axis as shown in Fig. \ref{Fig1_lattice}(b).
The magnetic anisotropy in the spin excitation energies is defined as
$\omega_n (\theta) = E_n (\theta)- E_0 (\theta)$, where $E_n$ and $E_0$ are the 
excited and ground state energy respectively. 
 This anisotropy is affected by all interactions other than the isotropic Heisenberg  limit ($J_{XY} = J_Z$), making it difficult to quantify the effect of individual interactions. However, if we compare the two excitation anisotropies, $\omega_n(\theta)$ and $\omega_n(-\theta)$ for a given strength $h$ and $\phi =0$ as shown in Fig. \ref{Fig1_lattice}(c), 
 related by $C_{2a}$ symmetry transformation, we can eliminate the effects of all other interactions except $J_{ac}$ thanks to symmetries of the model. 
 Since our theory relies on the symmetry of the Hamiltonian, the ground state should break the $C_{2a}$ symmetry only explicitly from the $J_{ac}$ term. The magnetic field also contributes to the $C_{2a}$ breaking, but by comparing two angles of $\theta$ and
$-\theta$, the effect of $J_{ac}$ is isolated.

We focus on the lowest energy excitation $n=1$ which gives a dominant resonance at low temperatures, and drop the $n$ in $\omega_n$ from now on for simplicity, even though our proposal works for all $n$.
  We define the excitation anisotropy between the magnetic field with angles of $\theta$ and $-\theta$ as
  $\delta\omega_{K} (\theta) \equiv \omega(\theta)-\omega(-\theta)$ for $0 < \theta < \frac{\pi}{2}$, 
  and the conventional anisotropy between in- and out-of-plane fields as $\delta \omega_A \equiv \omega({\theta = 0}) - \omega (\theta =\frac{\pi}{2})$.
  Below we first show how $\delta\omega_K$ arises from $J_{ac}$ under the field in the $a-c$ plane based on the symmetry. 
  
\noindent {\bf Symmetry Analysis} --  
To understand the origin of a finite $\delta\omega_{K}$ for $\phi=0$ under the magnetic field sweep, we first begin with a special case 
when $\phi = \frac{\pi}{2}$, i.e,  when the external field is in the $b-c$ plane. This is a special case where $\delta\omega_{K}=0$ for the following reason. 

The Zeeman terms due to the field with the angle $\theta$ and with $-\theta$ are related by a $\pi$ rotation of the field about the $\hat{b}$ axis, denoted by 
\begin{equation}
C_{2b,\theta}:\;
\mathcal{H}_{B}\propto(\cos\theta S_{i}^{b}+\sin\theta S_{i}^{c})\stackrel{}{\longrightarrow}
(\cos\theta S_{i}^{b}-\sin\theta S_{i}^{c}). 
\end{equation}
The same can be achieved by a $\pi$-rotation of the lattice,
\begin{equation}
C_{2b}:\; (S^{a},S^{b},S^{c})\rightarrow(-S^{a},S^{b},-S^{c}) \;\; {\rm and} \;\; \phi_{x}\leftrightarrow\phi_{y} , 
\end{equation}
which also indicates $\mathcal{H}$ is invariant under $C_{2b}$.
While $\mathcal{H}_{B}$ breaks the $C_{2b}$ symmetry of $\mathcal{H}$, the total Hamiltonian
 $\mathcal{H}+ \mathcal{H}_B (\theta)$ and $\mathcal{H}+ \mathcal{H}_B (-\theta)$ are related by $C_{2b}$ and therefore, share the same eigenenergies, i.e., $\delta\omega_K=0$. The difference due to the field is simply removed by
  a $\pi$ rotation of the eigenstates about the $\hat{b}$ axis.
The magnetic field sweeping from $\theta$ to $-\theta$ in the other planes equivalent to $b-c$ plane by $C_{3c}$ symmetry also gives $\delta\omega_{K}=0$.

Now let us consider when the magnetic field sweeps in the $a-c$ plane.  Similarly,  the magnetic field directions $\theta$ and $-\theta$ are
related by 
\begin{equation}
C_{2a,\theta}:
\mathcal{H}_{B}\propto(\cos\theta S_{i}^{a}+\sin\theta S_{i}^{c})\stackrel{}{\longrightarrow}(\cos\theta S_{i}^{a}-\sin\theta S_{i}^{c}).
\end{equation}
Considering a $\pi$ rotation of the lattice about the $\hat{a}$ axis,  
\begin{equation}
C_{2a}:\; (S^{a},S^{b},S^{c})\rightarrow(S^{a},-S^{b},-S^{c}) \;\; {\rm and} \;\; \phi_{x}\leftrightarrow\phi_{y} ,
\end{equation}
we find 
$J_{XY}$, $J_{Z}$, $J_{ab}$, terms are invariant under $C_{2a}$, while the $J_{ac}$ terms transform as 
\begin{equation}
C_{2a}: \; J_{ac}  \stackrel{}{\rightarrow} -J_{ac}.
\end{equation}
By the same argument, if $J_{ac}=0$, $\mathcal{H}$ is invariant under $C_{2a}$, and the eigenenergies of the total Hamiltonian for $\theta$ and $-\theta$ are the same, i.e., $\delta\omega_{K}=0$. If $J_{ac}\ne0$, the total Hamiltonian $\mathcal{H}+ \mathcal{H}_B (\theta)$ and $\mathcal{H}+ \mathcal{H}_B (-\theta)$ cannot be related by $C_{2a}$, and therefore, $\delta\omega_K\ne0$. We need to change the sign of $J_{ac}$ for the $C_{2a}$ relation to hold,  i.e.,
the transformation of the external field angles of $\theta$ to $-\theta$ is equivalent to the change of $J_{ac}$ to 
$-J_{ac}$.
Thus, the lack of $C_{2a}$ symmetry allows us to single out the $J_{ac}$ interaction 
through $\delta\omega_{K}$. 

Since $J_{ac}$ contains a combination of the Kitaev and $\Gamma$ interactions, we need other methods to
subtract the $\Gamma$ contribution.  The in- and out-of-plane anisotropy, $\delta \omega_A$ offers precisely the other information.
We note that the in- and out-of-plane anisotropy $\delta\omega_A$ is determined by $J_Z - J_{XY} = \Gamma$.
Thus, for the ideal edge sharing octahedral environment, we can first estimate $\Gamma$ from the measured $\delta \omega_A$, and then extract 
the Kitaev strength by subtracting the $\Gamma$ contribution from the measured $\delta \omega_K (\theta)$. 

Below we show numerical results of spin excitations obtained by ED on a 24-site cluster which can be measured by
angle-dependent FMR and INS techniques 
under magnetic field angles of $\theta$ and $-\theta$ with $\phi=0$. 

\begin{figure} 
\includegraphics[width=1.0\linewidth,trim={0mm 0mm 0mm 0mm}]{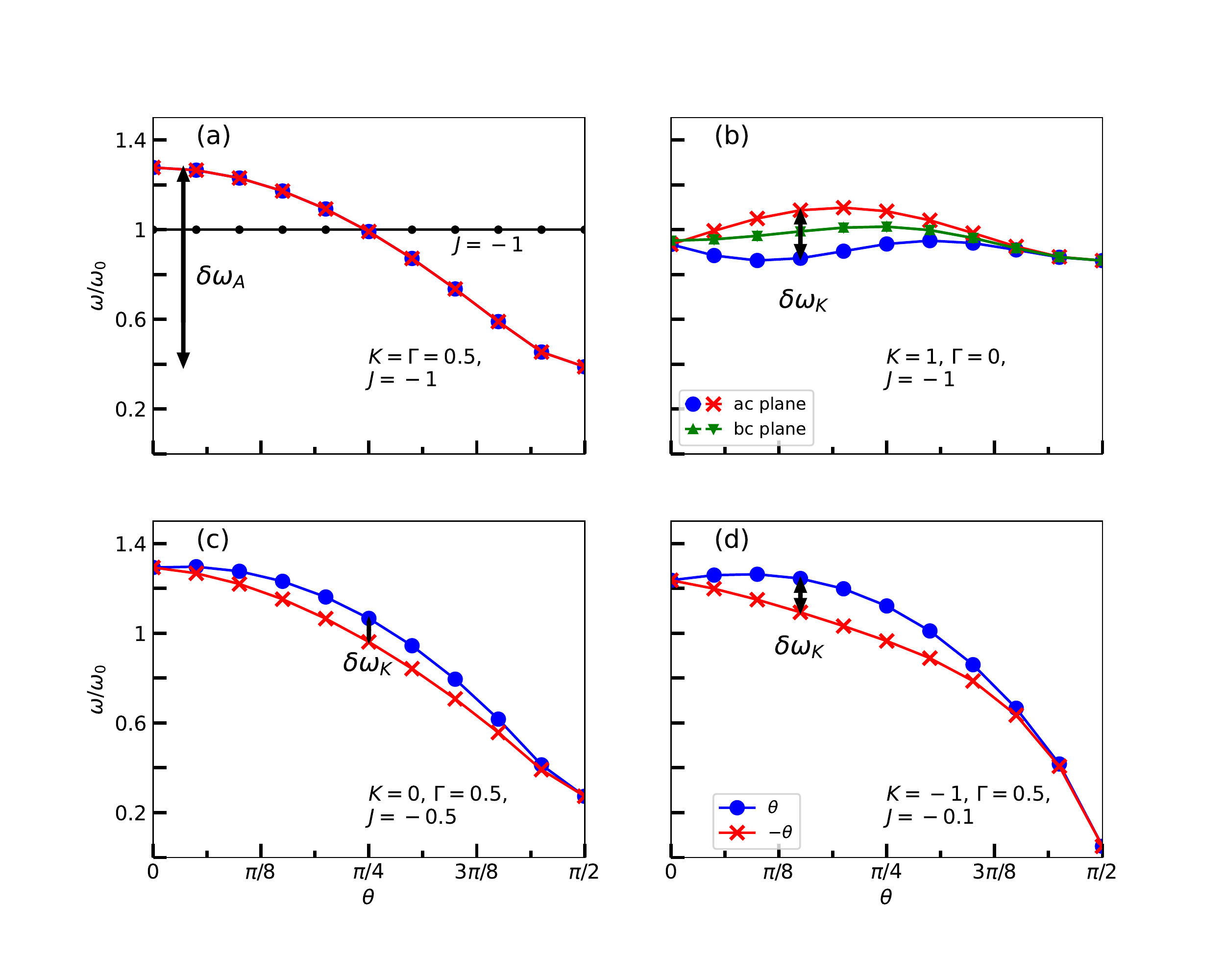}
\caption{\textbf{Angle-dependent spin excitations in ferromagnetic resonance (FMR) using exact diagonalizaiton on a $C_3$-symmetric 24-site cluster.} Various sets of parameters with Zeeman energy $g\mu_Bh=1$ are used. $\delta \omega_A$ is the difference in the spin excitation energies $\omega$ between fields along a-axis and c-axis, and $\delta \omega_K$ is the difference between $\omega(\theta)$ (blue) and $\omega(-\theta)$ (red), as highlighted by the arrows. $J$, $K$ and $\Gamma$ are the Heisenberg, Kitaev and off-diagonal interactions respectively. (a) $J=-1$ and $K=\Gamma=0.5$. (b) $J=-1$, $K=1$, and $\Gamma=0$. FMR in the $b-c$ plane is shown in green: $\theta$ (up triangle) and $-\theta$ (down triangle). (c) $J=-0.5$, $\Gamma=0.5$, and $K=0$. (d) $J=-0.1$, $K=-1$, $\Gamma=0.5$. See the FMR subsection for implication of the results.}
\label{Fig2_FMR}
 \end{figure}

\noindent{\bf Angle-Dependent Ferromagnetic Resonance} -- 
FMR is a powerful probe to study ferromagnetic or spin correlated materials. FMR spectrometers record the radio-frequency (RF) electromagnetic wave that is absorbed by the sample of interest placed under an external magnetic field. To observe the resonance signal, the resonant frequency of the sample is changed to match that of the RF wave under a scan of the external magnetic field, so the excitation anisotropy $\delta\omega(\theta)$ leads to the anisotropy in the resonant magnetic field. FMR provides highly resolved spectra over a large energy range and has been used to investigate exchange couplings \cite{Z1994, B2003,Nascimento2006,Lenz2003} and anisotropies \cite{DiazdeSihues2007, VG2006} due to its dependence on the magnetic field angle. 
Here, for simplicity, we calculate the excitation energy probed by the RF field (details can be found in the Methods) with a set magnetic field strength for spin $\frac{1}{2}$ using  ED on a $C_3$-symmetric 24-site cluster. 

We set our units the magnetic field $h = 1$ and $g=\mu_{B} \equiv1$, leading to the excitation energy of a free spin, $\omega_{0}=g\mu_B h=1$, so the excitation energies calculated are normalized by $\omega_{0}$.  
A few sets of different interaction parameters (in units of $\omega_0$) are investigated. 
Figure \ref{Fig2_FMR}(a) shows the $J=-1$ and $K=\Gamma=0.5$ case with no $\delta\omega_{K}(\theta)$ between $-\pi/2<\theta<0$ (red line) and $0<\theta<\pi/2$ (blue line), since $J_{ac}=0$. The conventional anisotropy $\delta\omega_{A}$ is finite, because the $\Gamma$ interaction generates a strong anisotropy between the plane $\theta=0$ and the c-axis $\theta=\pi/2$, i.e., $J_{XY}\neq J_{Z}$ due to a finite $\Gamma$ contribution. 
The black line is for only $J = -1$ showing a uniform FMR independent of angles which serves as a reference. 
Figure \ref{Fig2_FMR}(b) shows the $J=-1$, $K=1$, and $\Gamma =0$ case, which shows a finite $\delta\omega_{K}(\theta)$ between $-\pi/2<\theta<0$ and $0<\theta<\pi/2$ in the $a-c$ plane. On the other hand, no $\delta\omega_{K}(\theta)$ by sweeping $\theta$ in the $b-c$ plane (up and down triangles with green line) is observed, consistent with the symmetry analysis presented above. 
{Note the conventional anisotropy $\delta\omega_{A}$ in both $a-c$ and $b-c$ planes are not exactly zero,  because the Kitaev interaction selects the magnetic moment along the cubic axes in the ferromagnetic state via order by disorder\cite{GiniyatPRB2016,stavropoulos2018counter}. This leads to a tiny anisotropy between the plane $\theta=0$ and the c-axis $\theta=\pi/2$ when $\Gamma=0$ and $J_{XY}=J_{Z}$.  This anisotropy becomes weaker when the magnetic field increases, i.e, when the moment polarization overcomes the order by disorder effect. {\color{blue} Supplementary Note 1} shows that the anisotropy is almost gone when the field is increased by three times with the same set of parameters, where the Heisenberg limit (black line) is added for a reference. When $\Gamma$ becomes finite favouring either the $a-b$ plane or the c-axis depending on the sign of the $\Gamma$, this conventional anisotropy is determined by the $\Gamma$ interaction as shown in Fig. \ref{Fig2_FMR}(c) and (d), and the order by disorder effect becomes silent.}
Figure \ref{Fig2_FMR}(c) shows the $J=-0.5$, $\Gamma=0.5$, and $K=0$ case.
The $\Gamma$ interaction alone can generate a finite $\delta\omega_{K}$ due to the broken $C_{2a}$ by $J_{ac}$.
In addition, the $\Gamma$ interaction generates a large $\delta\omega_{A}$, different from Fig. \ref{Fig2_FMR}(b).
Figure \ref{Fig2_FMR}(d) presents the $J=-0.1$, $K=-1$ and $\Gamma = 0.5$ case, which is close to a set of parameters proposed for J$_{\rm eff} =\frac{1}{2}$ Kitaev candidate materials\cite{janssen2017model}. 
Clearly, $\delta\omega_{K}(\theta)$ is significant due to a finite $J_{ac}$, and $\delta\omega_{A}$ is also large due to a finite $\Gamma$. 
While a magnetic field of strength $h =1 $ is used to polarize the ground state where the finite-size effect is small as shown in {\color{blue} Supplementary Note 2},  our  symmetry argument works for any finite field.   
However,  we note that the finite-size effect of ED is minimal when the ground state is polarized.

\begin{figure} 
\includegraphics[width=0.9\linewidth,trim={0mm 0cm 0cm 0cm}]{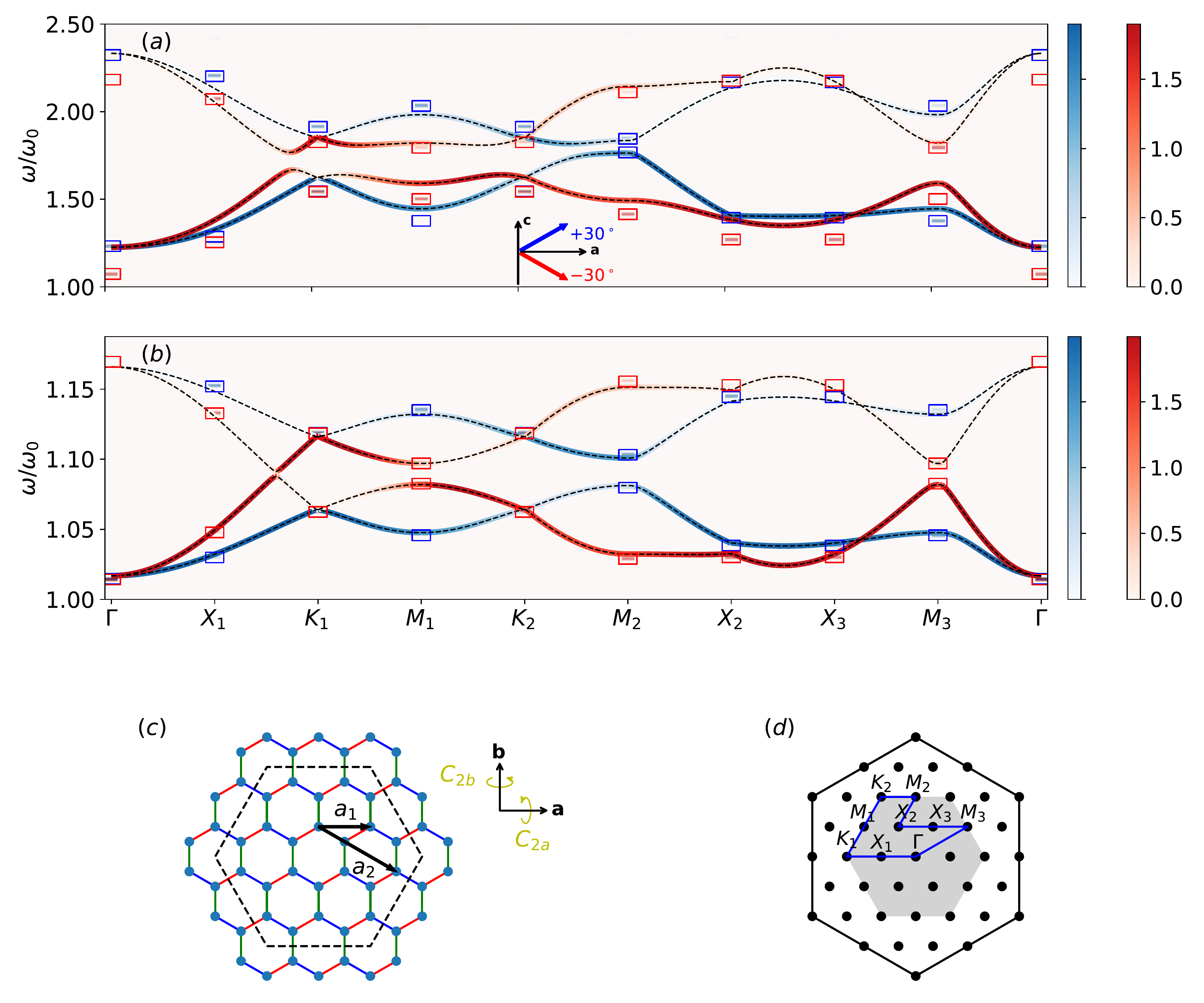}
\caption{ \textbf{Dynamic spin structure factor (DSSF) of the spin excitations at accessible wavevectors using exact diagonalizaiton (ED) on a $C_3$-symmetric 24-site cluster and linear spin wave theory (LSWT).}  The boxes and the dashed lines are DSSF obtained by ED and LSWT respectively. The colour bars represent the intensity of DSSF.  The same parameters for Fig. \ref{Fig2_FMR}(d) are used, i.e.  $(J, K, \Gamma) = (-0.1, -1, 0.5)$
in units of $\omega_0=g \mu_B h=1$. The magnetic field angles in the $a-c$ plane are $30^\circ$ (blue) and $-30^\circ$ (red). 
(b) DSSF with the same parameters as (a) except a larger field $g\mu_B h=8$, showing a better match between the ED and LSWT results; see the Inelastic Neutron Scattering subsection for further discussions. 
 (c) $C_3$-symmetric 24-site cluster used for the ED. (d) Accessible momentum points labeled in the x-axis of (a) and (b).}
\label{Fig3_INS}
 \end{figure}
\noindent{\bf Inelastic Neutron Scattering} --
Complementary to FMR, INS can measure excitations between different points in the reciprocal space based on the momentum transfer of the scattered neutrons. 
The magnon dispersions of the ordered states of magnetic materials measured via INS have been used to determine the spin exchange Hamiltonian parameters \cite{Shirane1968,Endoh1974,choi2012prl,banerjee2016proximate,Ran2017,Do2017,Banerjee2018excitations}.
Figure \ref{Fig3_INS}(a) and (b) show the spin excitations at accessible wavevectors on a $C_3$-symmetric 24-site cluster with the same exchange parameters for Fig. \ref{Fig2_FMR}(d)
and with $h=1$ and $h=8$, respectively. The cluster and the accessible momenta are shown in (c) and (d), respectively.
We set the magnetic field angles $\theta=30^\circ$ (blue) and $\theta=-30^\circ$ (red) in the $a-c$ plane. 
 The square boxes denote the excitation energies obtained by the ED, and the colour bars indicate the intensity of DSSF $\sum_{\alpha}S^{\alpha\alpha}(\boldsymbol{q},\omega)$ (details can be found in the Methods). The structure factor is convolved with a Gaussian of finite width to emulate finite experimental resolution. 
We observe a clear difference between the two field directions, $\delta\omega_K$ at every momentum points. 
In particular, $\delta\omega_K$ is the largest at $M_2$-point, while it is tiny at the $K_1$-point. Note that $M_1$ and $M_3$ are related by the $C_{2b}$ and inversion.

To gain more insights of $\delta\omega_{K}(\theta)$ at finite momenta obtained by ED, we also perform LSWT calculations 
 with the magnetization making an angle $\theta_M$ as indicated in Fig. \ref{Fig1_lattice}(b). $\theta_M$ is found via minimizing
the classical ground state energy (details can be found in the Methods); the LSWT with the set of parameters used for Fig. \ref{Fig3_INS}(a)\rq{}s ED results leads to $\theta_M \sim 12.1^\circ$. The 
spin excitations within the LSWT are shown as dashed lines together with the ED results
in Fig. \ref{Fig3_INS}(a).
The mismatch between LSWT and ED is visible at every momentum, which implies the significant effects of nonlinear terms\cite{Consoli2020PRB}.

However, when the field increases, the difference should decrease, since the magnetic polarization increases at a higher field.
In Fig. \ref{Fig3_INS}(b), we show both ED and LSWT with $h = 8$ and $\theta_M \sim 25.8^\circ$, where the two results match well as expected, and
the nonlinear terms become less significant.
In particular, the anisotropy $\delta \omega_K$ at the $K$-point at the high field limit given by the leading terms in $1/h$, is simplified as 
\begin{equation}
\begin{split}
\delta \omega_K (\theta) &= \frac{3}{8} \cos{\theta_M} \left( |2\sqrt{2}J_{ac} \sin{\theta_M} - J_{ab} \cos{\theta_M}|-  |2\sqrt{2}J_{ac} \sin{\theta_M} + J_{ab} \cos{\theta_M}|\right) \\
&\qquad+ \frac{9\sqrt{2}J_{ac}J_{ab}\left(2\sin2\theta_M+\sin4\theta_M\right)}{128h\cos(\theta-\theta_M)} + \mathcal{O}(\frac{1}{h^2}),
\label{Kpoint}
\end{split}
\end{equation}
where $\theta_M(\theta) \rightarrow \theta$ when $h \rightarrow \infty$.
This shows that both $J_{ac}$ and $J_{ab}$ should be finite for a finite $\delta\omega_K$ at the $K$-point,  which explains no splitting of $\delta\omega_K$ at the $K$-point in Fig. \ref{Fig3_INS}(b), as our choice of parameters gives $J_{ab} = 0$, i.e, $\Gamma = -K/2$. 
On the other hand, at the $M_2$-point, there is no simple expression,  but the leading terms of $\delta\omega_{K}(\theta)$ in $\delta\theta_{a/c}$ around the $a$- and $c$-axis ($\delta\theta_{a} = 0-\theta$ and $\delta\theta_c = \theta-\pi/2$) 
are given by
\begin{equation} \label{eq:expansion_M2}
\delta\omega_{K}(\theta)\simeq\begin{cases}
J_{ac}(\delta\theta_a)A + \mathcal{O}(\delta\theta_a^3)\\
J_{ac}(\delta\theta_c)C + \mathcal{O}(\delta\theta_c^3),
\end{cases}
\end{equation}
where $A$ and $C$ are functions of other interactions given in {\color{blue} Supplementary Note 3}. 
Clearly, $\delta\omega_{K}(\theta)$ appears as odd powers of $J_{ac}$ and $\delta\theta_{a/c}$, consistent with the symmetry analysis presented above.

So far, we have focused on the ideal octahedra environment. However, trigonal distortion is often present, albeit small, which introduces
extra exchange interactions. 
Below we discuss other contributions to $\delta \omega_A$ complicating the isolation of $K$ from $J_{ac}$ and our resolution of such complication in order to estimate the Kitaev interaction out of
a full Hamiltonian. 

\noindent{\bf Effects of trigonal distortion and further neighbour interactions} -- 
In principle, there are other small but finite interactions; few examples in $\delta \mathcal{H}^\prime$ include 
\begin{equation}\begin{split}
\delta \mathcal{H}^\prime &= \sum_{\langle ij\rangle\in\alpha\beta(\gamma)} \Big[\Gamma^{\prime}(S_{i}^{\alpha}S_{j}^{\gamma}+S_{i}^{\gamma}S_{j}^{\alpha}+S_{i}^{\beta}S_{j}^{\gamma}+S_{i}^{\gamma}S_{j}^{\beta})\Big]\\&\qquad+
J_2 \sum_{\langle\langle i,j\rangle\rangle} {\bf S}_i \cdot {\bf S}_j + 
J_3 \sum_{\langle\langle\langle i,j\rangle\rangle\rangle} {\bf S}_i \cdot {\bf S}_j,
\end{split}\label{deltaH}\end{equation}
where $\Gamma^\prime$ is introduced when a trigonal distortion is present\cite{rau2014trigonal}; 
 $J_2$ and $J_3$ are the second and third n.n. Heisenberg interactions respectively. 
It is natural to expect that they are smaller than the n.n. Kitaev, Gamma, and Heisenberg interactions\cite{HSKim2016structure,winter2016challenges,janssen2017model}.
Several types of interlayer exchange interactions are present,  but 
 they are even smaller than the terms considered in Eq. \ref{deltaH}\cite{HSKim2016structure}.
%

Let\rq{}s investigate how they affect the above analysis done for the ideal n.n. Hamiltonian. 
First of all, the isotropic interactions such as further neighbour $J_2$, $J_3$, and the interlayer Heisenberg do not make any change to our proposal, since they do not contribute to $\delta\omega_A$ nor $\delta\omega_K$.
On the other hand, the $\Gamma^\prime$ modifies the exchange parameters as follows: 
\begin{equation}\begin{split}
J_{XY}&=J+J_{ac}-\Gamma^{\prime},\;\; J_{Z}=J+J_{ab}+2\Gamma^{\prime},
\\J_{ab}&=\frac{1}{3}K+\frac{2}{3}(\Gamma-\Gamma^{\prime}), \;\; J_{ac}=\frac{1}{3}K-\frac{1}{3}(\Gamma-\Gamma^{\prime}).
\end{split}
\label{relationJKG}
\end{equation}
The conventional anisotropy $\delta\omega_A$ is now due to $\Gamma + 2 \Gamma^\prime$ obtained from $J_{Z} - J_{XY}$.
Thus to single out the Kitaev interaction, one has to find both $\Gamma$ and $\Gamma^\prime$, as $J_{ac}$ is a combination of $K$, $\Gamma$ and $\Gamma\rq{}$. 
 Once the trigonal distortion is present, the g-factor also becomes anisotropic, i.e., the in-plane $g_{a}$ is different from
the c-axis $g_c$, which affects $\delta\omega_A$.

However, the g-factor anisotropy does not affect the $\delta\omega_K$, since the field angles of $\theta$ and $-\theta$ involve
the same strength of in- and out-of-plane field components, i.e, ${\bf h} (\theta) = h_a {\hat a} + h_c {\hat c}$ and 
${\bf h} (-\theta) = - h_a {\hat a} + h_c {\hat c}$.  Thus  we wish to extract the information of $K$ and $\Gamma - \Gamma^\prime$
from $\delta \omega_K$,  as it is free from the g-factor anisotropy.

We note that $\delta\omega_K$ at the K-point, Eq. (\ref{Kpoint}) offers both $J_{ac}$ and $J_{ab}$ from 
the first term independent of the field and the next term proportional to $1/h_{\rm eff}$ ($h_{\rm eff}=h\sqrt{g_{a}^2\cos^2\theta + g_c^2\sin^2\theta}$). 
Once $J_{ac}$ and $J_{ab}$ are deduced, $K$ and $\Gamma - \Gamma\rq{}$ can be estimated from Eq. (\ref{relationJKG}).
The measurements of $\delta\omega_{K}$ at the $K$-point with a large magnetic field then determine $K$ and $\Gamma - \Gamma^\prime$ separately.
Further neighbor Heisenberg interactions, $J_2$ and $J_3$ do not modify Eq. (\ref{Kpoint}) in the high-field limit, so they do not affect our procedure.

\vspace{-2mm}\section*{\hspace{-7mm} Discussion}\vspace{-2mm}

We propose an experimental setup to single out the Kitaev interaction for  honeycomb Mott insulators with edge-sharing octahedra.
In an ideal octahedra cage, the symmetry-allowed  n.n. interactions contain the Kitaev, another bond-dependent $\Gamma$ and Heisenberg interactions.
We prove that the magnetic anisotropy related by the $\pi$-rotation around the $a$-axis denoted by $\delta\omega_K$
occurs only when a combination of $K$ and $\Gamma$,
i.e. $K-\Gamma$, is finite.  This can be measured from the spin excitation energy differences under the magnetic field of angle sweeping from above to below the honeycomb plane using the FMR or INS techniques. 
Since the in- and out-of-plane magnetic anisotropy, $\delta\omega_A$ is determined solely by $\Gamma$, one can estimate $\Gamma$ strength first from $\delta\omega_A$ and then extract the Kitaev interaction from
$\delta\omega_K$.

While the trigonal distortion introduces an additional interaction, the Kitaev interaction is unique
as it is the only interaction that contributes to $\delta\omega_K$ without altering $\delta\omega_A$. 
Our theory is applicable to all Kitaev candidate materials including an emerging candidate RuCl$_3$. 
In particular, since the two dominant interactions are 
ferromagnetic Kitaev and positive $\Gamma$ interactions in RuCl$_3$ \cite{HSKim2016structure,winter2016challenges,winter2017review,Janssen2019review}, leading to a large $J_{ac}$ and a small $J_{ab}$, we predict that 
$\delta\omega_K$ independent of the g-factor anisotropy is significant except at the $K$-point. 
{\color{blue} Supplementary Note 4} shows the FMR and INS of a set of parameters with a small negative $\Gamma^\prime$ interaction to stabilize a zero-field zig-zag ground state as in RuCl$_3$ \cite{HSKim2016structure,winter2016challenges,janssen2017model}.
 Another relevant perturbation in some materials is the effect of monoclinic structure which loses the $C_{3c}$ symmetry of $R{\bar 3}$, making the z-bond different from the x- and y-bonds.
The current theory of finite $\delta\omega_K$ due to a finite $J_{ac}$ still works for $C2/m$ structure. 
However, since the z-bond of $J_{ac}^z (=K_z/3-\Gamma_z/3)$ is no longer the same as the x- and y-bonds
of $J_{ac}^{x} (= J_{ac}^y)$ and $C_{2a}$ symmetry relates between the x- and y-bonds, the anisotropy $\delta \omega_K$ at different momenta, detecting both $J_{ac}^{x} = K_x/3 - \Gamma_x/3$ and $J_{ac}^{z}$, is required to determine different x- and z-bond strengths.

    The symmetry-based theory presented here is also valid for higher spin models with the Kitaev interaction such as $S=3/2$ CrI$_3$ including a nonzero single-ion anisotropy\cite{Xu2018Kitaev,Peter2019HigherK,Lee2020PRL,StavropoulosPRR2021} which generates a further anisotropy in $\delta\omega_{A}$ but does not affect the $\delta \omega_K$.  The next nearest neighbor Dzyaloshinskii-Moriya interaction with the d-vector along the c-axis\cite{ChenPRX2021} is also invariant under the $C_{2a}$ symmetry.  Further studies for higher-spin models remain to be investigated to identify higher-spin Kitaev spin liquid.
We would like to emphasize that the proposed set-up is suitable for other experimental techniques such as low-energy terahertz optical and nuclear magnetic resonance spectroscopies that probe spin excitations in addition to the angle-dependent FMR and INS spectroscopy shown in this work as examples.

\vspace{-2mm}\section*{\hspace{-7mm} Methods}\vspace{-2mm}

\noindent {\bf Exact Diagonalization Simulations} -- Numerical ED was used to compute spin excitations under a magnetic field. ED was performed on a 24-site honeycomb cluster with periodic boundary conditions, where the Lanczos method \cite{Lanczos1950, Weie2008} was used to obtain the lowest-lying eigenvalues and eigenvectors of the Hamiltonian in Eq.~(\ref{eq:hamiltonian}).
The 24-site honeycomb shape and accessible momentum points in the Brillouin zone are shown in the Fig. \ref{Fig3_INS}(c) and (d).
The probability of the spin excitation of momentum $\boldsymbol{q}$ and energy $\omega$ is proportional to the dynamic spin structure factor \cite{Lovesey1984}(DSSF) given by
\begin{equation}\begin{split}
S^{\alpha\beta}(\boldsymbol{q},\omega)&=\frac{1}{N}\sum_{i,j}^{N}e^{-i\boldsymbol{q}\cdot(\boldsymbol{R}_{i}-\boldsymbol{R}_{j})}\int_{\infty}^{\infty}dte^{i\omega t}\left\langle S_{i}^{\alpha}(t)S_{j}^{\beta}(0)\right\rangle \\&=\frac{1}{N}\sum_{i,j}^{N}e^{-i\boldsymbol{q}\cdot(\boldsymbol{R}_{i}-\boldsymbol{R}_{j})}\sum_{\lambda,\lambda^{\prime}}p_{\lambda}\langle\lambda|S_{i}^{\alpha}|\lambda^{\prime}\rangle\langle\lambda^{\prime}|S_{j}^{\beta}|\lambda\rangle\delta(\hbar\omega+E_{\lambda}-E_{\lambda^{\prime}})\\&=\sum_{\lambda,\lambda^{\prime}}p_{\lambda}\langle\lambda|S_{-\boldsymbol{q}}^{\alpha}|\lambda^{\prime}\rangle\langle\lambda^{\prime}|S_{\boldsymbol{q}}^{\beta}|\lambda\rangle\delta(\hbar\omega+E_{\lambda}-E_{\lambda^{\prime}}),
\end{split}\end{equation}
 where the Lehmann representation is used; $|\lambda\rangle$ and $|\lambda^{\prime}\rangle$ are the eigenstates with the thermal population factor
 $p_{\lambda}$, and $S^{\alpha,\beta}$ are the spin operators. 
In the low temperatures, we take $|\lambda\rangle$ to be the ground state $|0\rangle$ and we are interested in the lowest energy excitation to $|1\rangle$ with a nonzero probability. 
For optical spectroscopies such as FMR, $\alpha=\beta=$ direction of the RF electromagnetic field and $\boldsymbol{q}=0$, so $|0\rangle$ and $|1\rangle$ belong to the same momentum sector. The structure factor simplifies to
$$
S^{\alpha\alpha}(\omega)=\frac{1}{N}\left|\langle1|\sum_{i}S_{i}^{\alpha}|0\rangle\right|^{2}\delta(\hbar\omega+E_{0}-E_{1}).
$$
For INS, the finite $\boldsymbol{q}$ must match the difference in the momenta of $|0\rangle$ and $|1\rangle$. For simplicity, we calculate the DSSF for $\alpha=\beta$,
$$
\sum_{\alpha}S^{\alpha\alpha}({\bf q}, \omega)=\sum_{\alpha}\left|\langle1|S_{\boldsymbol{q}}^{\alpha}|0\rangle\right|^{2}\delta(\hbar\omega+E_{0}-E_{1}).
$$


\noindent{\bf Linear Spin Wave Theory} -- 
The Hamiltonian in Eq.~(\ref{eq:hamiltonian}) is bosonized by the standard Holstein-Primakoff transformation \cite{Holstein1940} expanded to linear order in the spin $S$:
\begin{equation}\begin{split}
S_{j}^{+}&=S_{j}^{a}+iS_{j}^{b}=\sqrt{2S}\left(a_{j}-\frac{a_{j}^{\dagger}a_{j}a_{j}}{4S}+\mathcal{O}(\frac{1}{S^{2}})\right)\simeq\sqrt{2S}a_{j}\\S_{j}^{-}&=S_{j}^{a}-iS_{j}^{b}=\sqrt{2S}\left(a_{j}^{\dagger}-\frac{a_{j}^{\dagger}a_{j}^{\dagger}a_{j}}{4S}+\mathcal{O}(\frac{1}{S^{2}})\right)\simeq\sqrt{2S}a_{j}^{\dagger}\\S_{j}^{c}&=S^{c}-a_{j}^{\dagger}a_{j},
\end{split}\end{equation}
where the quantization axis is parallel to the c-axis.
 The Fourier transforms are $a_{j}=\frac{1}{\sqrt{N}}\sum_{\boldsymbol{k}}e^{i\boldsymbol{k}\cdot\boldsymbol{r}_{j}}a_{\boldsymbol{k}}$ for sublattice A and
 $b_{j}=\frac{1}{\sqrt{N}}\sum_{\boldsymbol{k}}e^{i\boldsymbol{k}\cdot(\boldsymbol{r}_{j}+\delta)}b_{\boldsymbol{k}}$ for sublattice B, where $\delta$ is the vector pointing to nearest neighbors. The resulting quadratic Hamiltonian has the form $\mathcal{H}=\sum_{\boldsymbol{k}}\text{X}^{\dagger}H(\boldsymbol{k})\text{X}$, where $\text{X}^{\dagger}=(\begin{array}{cccc}
a_{\boldsymbol{k}}^{\dagger}, & b_{\boldsymbol{k}}^{\dagger}, & a_{-\boldsymbol{k}}, & b_{-\boldsymbol{k}}\end{array})$. 
Diagonalizing this BdG Hamiltonian following standard methods \cite{Colpa1978} gives two spin wave excitation branches.

For a general field, the Hamiltonian in Eq.~(\ref{eq:hamiltonian}) is first written in new axes $\ensuremath{\boldsymbol{a}^{\prime}\boldsymbol{b}^{\prime}\boldsymbol{c}^{\prime}}$. 
$\boldsymbol{a}^{\prime}=(\sin\theta_M\cos\phi_M,\sin\theta_M\sin\phi_M,-\cos\theta_M)$, $\boldsymbol{b}^{\prime}=(-\sin\phi_M,\cos\phi_M,0)$ and
$\qquad \qquad$  $\boldsymbol{c}^{\prime}=(\cos\theta_M\cos\phi_M,\cos\theta_M\sin\phi_M,\sin\theta_M)$. $\boldsymbol{c}^{\prime}$ is parallel to the magnetization $\bold{S}(\theta_M,\phi_M)$, which is not the same direction as the magnetic field, unless
the field is very large to fully polarize the moment.
The magnetization angles $(\theta_M,\phi_M)$ are obtained by minimizing the classical ground state energy,
and LSWT is applied on the ground state \cite{Consoli2020PRB}.
Arbitrary $\boldsymbol{a}^{\prime}$ and $\boldsymbol{b}^{\prime}$ axes obtained by rotation around $\boldsymbol{c}^{\prime}$ are valid and do not affect the result.

\vspace{-2mm}\section*{\hspace{-7mm} Data availability}\vspace{-2mm}
The data that support the findings of this study are available from the corresponding author upon reasonable request.

\vspace{-2mm}\section*{\hspace{-7mm} Code availability}\vspace{-2mm}
The code used to generate the data used in this study is available from the corresponding author upon reasonable request.

\bibliography{references3}
\vspace{2mm}

\noindent{\bf Acknowledgements}

\noindent We thank J. Gordon,  I. Lee, C. Hammel, S. Nagler, and A. Tennant for useful discussions.
 This work was supported by the Natural Sciences and Engineering Research Council of Canada and the Canada Research Chairs Program.
This research was enabled in part by support provided by Sharcnet (\href{http://www.sharcnet.ca}{www.sharcnet.ca}) and Compute Canada (\href{http://www.computecanada.ca}{www.computecanada.ca}).
Computations were performed on the GPC and Niagara supercomputers at the SciNet HPC Consortium. 
SciNet is funded by: the Canada Foundation for Innovation under the auspices of Compute Canada; the Government of Ontario; Ontario Research Fund - Research Excellence; and the University of Toronto.

\noindent{\bf Author Contributions}

\noindent Exact diagonalization and linear spin wave theory calculations were performed by J. C. The symmetry analysis was done by H. -Y. K. and J. C.
H.-Y.K. planned and supervised the project. All authors wrote the manuscript. 

\noindent{\bf Additional Information}

\noindent {\bf Supplementary Information} --  is available in the online version of the paper. 

\noindent {\bf Competing Interests} -- The authors declare no competing interests. Hae-Young Kee is an Editorial Board Member for Communications Physics, but was not involved in the editorial review of, or the decision to publish this article.

\noindent {\bf Correspondence} -- should be addressed to H.-Y.K. (\href{mailto:hykee@physics.utoronto.ca}{hykee@physics.utoronto.ca}).

\end{document}